\documentclass[aps,twocolumn,prl]{revtex4}
\usepackage{graphicx}
\usepackage{bm}

\begin{document}
\newcommand{\dx}{\delta\bm x}
\newcommand{\du}{\delta\bm u}
\newcommand{\da}{\delta\bm a}
\title{Timescales of Turbulent Relative Dispersion}

\author{Rehab Bitane} \affiliation{Laboratoire Lagrange, UMR7293,
  Universit\'e de Nice Sophia-Antipolis, CNRS, Observatoire de la
  C\^ote d'Azur, BP 4229, 06304 Nice Cedex 4, France}
\author{J\'er\'emie Bec} \affiliation{Laboratoire Lagrange, UMR7293,
  Universit\'e de Nice Sophia-Antipolis, CNRS, Observatoire de la
  C\^ote d'Azur, BP 4229, 06304 Nice Cedex 4, France} \author{Holger
  Homann} \affiliation{Laboratoire Lagrange, UMR7293, Universit\'e de
  Nice Sophia-Antipolis, CNRS, Observatoire de la C\^ote d'Azur, BP
  4229, 06304 Nice Cedex 4, France}
\begin{abstract}
  Tracers in a turbulent flow separate according to the celebrated
  $t^{3/2}$ Richardson--Obukhov law, which is usually explained by a
  scale-dependent effective diffusivity.  Here, supported by
  state-of-the-art numerics, we revisit this argument. The Lagrangian
  correlation time of velocity differences is found to increase too
  quickly for validating this approach, but acceleration differences
  decorrelate on dissipative timescales. This results in an asymptotic
  diffusion $\propto t^{1/2}$ of velocity differences, so that the
  long-time behavior of distances is that of the integral of Brownian
  motion. The time of convergence to this regime is shown to be that
  of deviations from Batchelor's initial ballistic regime, given
  by a scale-dependent energy dissipation time rather than the
  usual turnover time. It is finally argued that the fluid flow
  intermittency should not affect this long-time behavior of relative
  motion.
\end{abstract}
\maketitle

\noindent Turbulence has the feature of strongly enhancing the
dispersion and mixing of the species it transports. It is known since
the work of Richardson \cite{richardson:1926} that tracer particles
separate in an explosive manner $\propto t^{3/2}$ that is much faster
and less predictable than in any chaotic system. While little doubt
remains about its validity in three-dimensional homogeneous isotropic
turbulence, observations of this law in numerics and experiments are
difficult, as they require a huge scale separation between the
dissipative lengths, the initial separation of tracers, the
observation range and the integral scale of the flow
\cite{sawford:2001,salazar-collins:2009}.  Much effort has been
devoted to test the universality of this law, which was actually
retrieved in various turbulent settings, such as the two-dimensional
inverse cascade~\cite{jullien-etal:1999}, buoyancy-driven
flows~\cite{schumacher:2008}, and
magneto-hydrodynamics~\cite{busse-mueller:2008}. At the same time,
breakthroughs on transport by time-uncorrelated scale-invariant flows
have strenghtened the original idea of Richardson that this law
originates from the diffusion of tracer separation in a
scale-dependent environment~\cite{falkovich-etal:2001}.  As a result,
the physical mechanisms leading to Richardson--Obukhov $t^{3/2}$ law
are still rather poorly understood and many questions remain open on
the nature of subleading terms, the rate of convergence and on the
effects of the intermittent nature of turbulent velocity
fluctuations~\cite{boffetta-sokolov:2002b,biferale-etal:2005}.

Turbulent relative dispersion consists in understanding the evolution
of the separation $\dx(t) \!=\! \bm X_1(t) \!-\! \bm X_2(t)$ between
two tracers.  Richardson's argument can be reinterpreted by assuming
that the velocity difference $\du(t) \!=\! \bm u(\bm X_1,t) \!-\! \bm
u(\bm X_2,t)$ has a short correlation time. This means that the
central-limit theorem applies and that, for sufficiently large
timescales,
\begin{equation}
  \frac{\mathrm{d}\dx}{\mathrm{d}t} = \du \simeq
  \sqrt{\tau_L}\,\, \bm U(\dx) \,\bm \xi(t),
  \label{eq:diffsep}
\end{equation}
where $\bm \xi$ is the standard three-dimensional white noise, $\bm
U^\mathsf{T}\bm U \!=\!\langle\du \otimes\du\rangle$ the Eulerian
velocity difference correlation tensor, and $\tau_L$ the Lagrangian
correlation time of velocity differences between pair separated by
$\delta x\! =\! |\dx|$. As stressed by Obukhov \cite{obukhov:1941},
when assuming Kolmogorov 1941 scaling, $\tau_L\!\sim\! \delta
x^{2/3}$, $\bm U \!\sim\! \delta x^{1/3}$, and the Fokker--Planck
equation associated to (\ref{eq:diffsep}) exactly corresponds to that
derived by Richardson for the probability density $p(\delta x,t)$. It
predicts in particular that the squared distance $\langle
|\dx(t)|^2\rangle_{r_0}$ averaged over all pairs that are initially at
a distance $|\dx(0)|\!=\!r_0$ has a long-time behavior $\propto t^3$
that is independent on $r_0$. This memory lost on the initial
separation can only occur on time scales longer than the correlation
time $\tau_L(r_0)\!\sim\! r_0^{2/3}$ of the initial velocity
difference. For times $t\ll\tau_L(r_0)$, one cannot make use of the
approximation (\ref{eq:diffsep}) as the velocity difference almost
keeps its initial value. This corresponds to the ballistic regime
$\langle |\dx(t)\!-\!\dx(0)|^2\rangle_{r_0} \!\simeq \! t^2 S_2(r_0)$,
where $S_2(r) \!= \!\langle |\du|^2\rangle$ is the Eulerian
second-order structure function over a separation $r$, introduced by
Batchelor \cite{batchelor:1950}.  The diffusive approach
(\ref{eq:diffsep}) can however be modified to account for the
ballisitic regime \cite{kraichnan:1966}.  Nevertheless a
short-time correlation of velocity differences can hardly been derived
from first principles and seems to contradict turbulence
phenomenology. Indeed, as stressed in \cite{falkovich-etal:2001}, if
$\delta x$ grows like $t^{3/2}$, the Lagrangian correlation time
$\tau_{L}$ is of the order of $\delta x^{2/3} \!\sim \!t$, so that the
velocity difference correlation time is always of the order of the
observation time.  Despite such apparent contradictions, Richardson
diffusive approach might be relevant to describe some intermediate
regime valid for large-enough times and typical separations.  Several
measurements show that the separations distribute with a probability
that is fairly close to that obtained from an eddy-diffusivity
approach~\cite{biferale-etal:2005,ouellette-etal:2006,eyink:2011}.

To clarify when and where Richardson's approach might be valid, it is
important to understand the timescale of convergence to the explosive
$t^3$ law. Much work has recently been devoted to this issue: it was
for instance proposed to make use of fractional diffusion with
memory~\cite{ilyin-etal:2010}, to introduce random delay times of
convergence to Richardson scaling~\cite{rast-pinton:2011}, or to
estimate the influence of extreme events in particle
separation~\cite{scatamacchia-etal:2012}. All these approaches
consider as granted that the final behavior of separations is
diffusive. As we will see here, many aspects of the convergence to
Richardson's law for pair dispersion can be clarified in terms of a
diffusive behavior of velocity differences.

To address such issues, we make use of direct numerical
simulations. For this, the Navier--Stokes equation with a
large-scale-forcing is integrated in a periodic domain using a
massively parallel spectral solver at two different
resolutions. Table~\ref{table} summarizes the parameters of the
simulations (see \cite{grauer-etal:2010} for more details). In each
case, the flow is seeded with $10^7$ Lagrangian tracers. Their
positions, velocities, and accelerations are then stored with enough
frequency to study relative motion.
\begin{table}[h]
  \begin{tabular}{|c|c|c|c|c|c|c|c|c|} \hline
    $N$ & $R_\lambda$ & $\nu$ & $\epsilon$ & $u_\mathrm{rms}$ & $\eta$ &
    $\tau_\eta$ & $L$ & $T$\\ \hline
    $2048^3$ & $460$ & $2.5\!\cdot\!10^{-5}$ &  $3.6\!\cdot\!10^{-3}$ & $0.19$ &
    $1.4\!\cdot\!10^{-3}$ & $0.083$ & $1.85$ & $9.9$ \\\hline
    $4096^3$ & $730$ & $1.0\!\cdot\!10^{-5}$ &  $3.8\!\cdot\!10^{-3}$ & $0.19$ &
    $7.2\!\cdot\!10^{-4}$ & $0.05$ & $1.85$ & $9.6$ \\\hline
  \end{tabular}
  \caption{\label{table}
    Parameters of the numerical simulations. $N$ is the number
    of grid points, $R_\lambda$ the Taylor-based Reynolds number,
    $\nu$ the kinematic viscosity, $\epsilon$ the averaged energy
    dissipation rate, $u_\mathrm{rms}$ the root-mean square velocity,
    $\eta \!=\! (\nu^3/\epsilon)^{1/4}$ the Kolmogorov dissipative scale,
    $\tau_\eta \!=\!  (\nu/\epsilon)^{1/2}$ the associated turnover time,
    $L \!=\! u_\mathrm{rms}^3/\epsilon$ the integral scale and $T \!=\!
    L/u_\mathrm{rms}$ the associated large-scale turnover time.}
\end{table}

\begin{figure}[t]
  \includegraphics[width=\columnwidth]{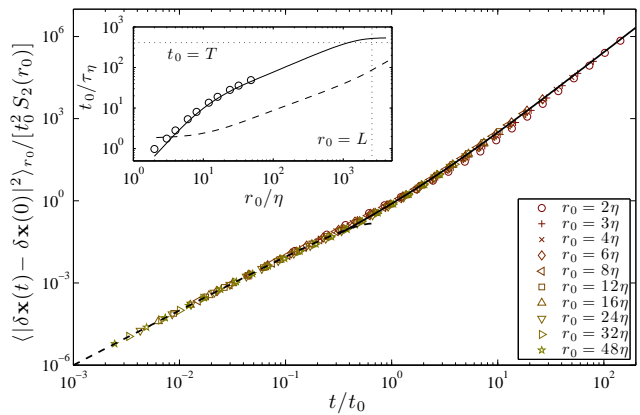}
  \vspace{-20pt}
  \caption{Time-evolution of the mean-square separation for
    $R_\lambda\!=\!730$ and different initial separations. The dashed
    line represents the two leading terms of the ballisitic behavior
    (\ref{eq:sep_smalltime}). The solid line is a fit to the
    Richardson regime (\ref{eq:sep_largetime}) with $g\!=\!0.52$ and
    $C\!=\!1.6$. Inset: $t_0$ as a function of $r_0$ in
    dissipative-scale units. The solid line is an Eulerian average,
    the circles are Lagrangian measurements and the dashed line is the
    turnover time $\tau(r_0)$.}
  \label{fig:batch_rich}
\end{figure}

We first report results on the behavior of the separation $\dx(t)$ as
a function of time. Following~\cite{ouellette-etal:2006}, a
Taylor expansion at short times leads to
\begin{equation}
  \left\langle |\dx(t)\!-\!\dx(0)|^2 \right\rangle_{r_0} = t^2 S_2(r_0)
  \!+\! t^3 \left\langle\du\cdot\da\right\rangle  \!+\! \mathrm{O}(t^4)\,,
\label{eq:sep_smalltime}
\end{equation}
where $S_2(r) = \langle |\du|^2\rangle$ is the second-order structure
function, $\langle\cdot\rangle$ denote Eulerian averages, and $\da(t)
\!=\! \bm a(\bm X_1,t) \!-\! \bm a(\bm X_2,t)$ is the difference of
the fluid acceleration sampled by the two tracers (where the notation
$\bm a \!=\! \partial_t \bm u \!+\! \bm u\cdot\nabla\bm u$).  As long
as the term $\propto t^2$ is dominant, the tracers separate
ballistically. Expansion (\ref{eq:sep_smalltime}) clearly fails for
$t\!\approx\! t_0 \!=\! S_2(r_0) /
|\left\langle\du\cdot\da\right\rangle |$.  It is known
\cite{mann-ott:2000,falkovich-etal:2001} that for separations in the
inertial range $\left\langle\du\cdot\da\right\rangle = -2\epsilon$,
which is nothing but a Lagrangian version of the $4/5$ law. This
implies that the ballistic regime ends up at times of the order of
\begin{equation}
  t_0 = {S_2(r_0)}/{(2\epsilon)}.
  \label{eq:deft0}
\end{equation}
This timescale can be interpreted as the time required to dissipate
the kinetic energy contained at the scale $r_0$. We thus expect it to
be equal to the correlation time of the initial velocity
difference. $t_0$ differs from the turnover time $\tau(r_0) =
r_0/[S_2(r_0)]^{1/2}$ defined as the ratio between the separation
$r_0$ and the typical turbulent velocity at that scale. When
Kolmogorov 1941 scaling is assumed, these two time scales have the
same dependency on $r_0$. However, usual estimates of the Kolmogorov
constant lead to $t_0/\tau(r_0) \approx 20$.  Also, note that
intermittency corrections to the scaling behavior of $S_2$ should in
principle decrease this ratio. Figure \ref{fig:batch_rich} represents
the mean-squared displacement rescaled by $t_0^2 S_2(r_0)$ as a
function of $t/t_0$, for various values of the initial separation
$r_0$. In such units and when $r_0$ is far in the inertial range, all
measurements collapse onto a single curve. The subleading term
$\propto t^3$ in~(\ref{eq:sep_smalltime}) is relevant for times
$t\lesssim0.01\,t_0$.

The data collapse extends to times larger than $t_0$ when the mean
squared separation tends to Richardson $t^3$ regime. This unexpected
fact implies that $t_0$ is not only the timescale of departure from
the ballistic regime, but also that of convergence to Richardson's
law. More precisely, numerical data suggest that for $t\gg t_0$
\begin{equation}
  \left\langle |\dx(t)\!-\!\dx(0)|^2 \right\rangle_{r_0} =
  g\,\epsilon\,t^3\left [ 1+ C({t_0}/{t})\right] + \mathrm{h.o.t.}.
  \label{eq:sep_largetime}
\end{equation}
The constant $C$ does not strongly depend on the Reynolds number.
Systematic measurements as a function of the initial separation show
that $C$ is negative when $r_0$ is of the order of the Kolmogorov
scale $\eta$. The convergence to Richardson law is then from below and
is thus contaminated by tracer pairs which spend long times close
together before sampling the inertial range; this is consistent with
the findings of \cite{scatamacchia-etal:2012}. When $r_0$ is
far-enough in the inertial range, $C\approx 1.6$ becomes independent
on the initial separation and the convergence to Richardson law is
from above. One finds that $C=0$ for $r_0\approx 4\eta$; the only
subleading terms in (\ref{eq:sep_largetime}) are then of higher order,
so that the mean-squared separation converges faster to Richardson
regime. Such an initial separation could be an ``optimal choice'' to
observe the $t^3$ behavior in experimental settings.

To understand why the timescale of convergence to Richardson law is of
the order of $t_0$, let us examine the timescales entering the
relative dispersion process. As already stated, the velocity
difference $\du$ between the two tracers stays correlated over a time
that increases too fast with the separation, making difficult to
justify the diffusive approach (\ref{eq:diffsep}). However, it is
known that turbulent acceleration, which is a small-scale quantity, is
correlated over times that are of the order of $\tau_\eta$ the
Kolmogorov turnover time \cite{mordant-etal:2004}. Its amplitude is
rather correlated on times of the order of the forcing correlation
time, but this does not alter the argument below.
\begin{figure}[h]
  \includegraphics[width=\columnwidth]{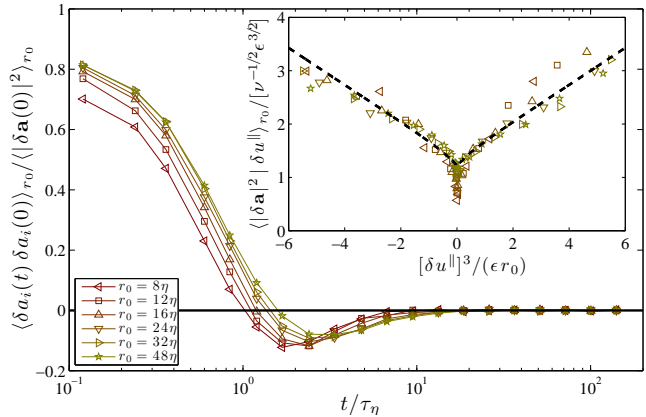}
  \caption{Lagrangian time autocorrelation of the acceleration
    difference $\da$ for various $r_0$ and
    $R_\lambda=730$. Inset: for the same separations $r_0$, variance
    of the acceleration difference amplitude conditioned on the
    longitudinal velocity difference $\delta u^\parallel$ as a
    function of the local dissipation rate $[\delta
    u^\parallel]^3/r_0$. }
  \label{fig:autocor_accel}
\end{figure}
Figure~\ref{fig:autocor_accel} represents the Lagrangian
autocorrelation of the difference of acceleration $\da$ between two
tracers. We clearly see that the components of this quantity
decorrelate on times of the order of $\tau_\eta$. This suggests
applying the central-limit theorem, so that
for separations in the inertial range and on timescales much longer
than the $\tau_\eta$, the difference of acceleration between two
tracers can be approximated by a delta-correlated-in-time random
process. We thus have
\begin{equation}
  \frac{\mathrm{d}\dx}{\mathrm{d}t} \!=\! \du, \mbox{ with
  }\frac{\mathrm{d}\du}{\mathrm{d}t} \!=\! \da \!\simeq\!
  \sqrt{\tau_\eta^\mathrm{loc}}\,\bm A(\dx,\du)\,\bm\xi(t),
 \label{eq:model_da}
\end{equation}
where $\bm A$ is defined as $\bm A^\mathsf{T}\bm A = \left\langle
  \da\otimes\da \,|\, \dx,\du\right\rangle$, $\bm\xi$ is the
three-dimensional white noise, and the product is here understood in
the Stratonovich sense.  The idea of assuming uncorrelated
accelerations is common to many stochastic models for turbulent
dispersion (see, e.g.,
\cite{kurbanmuradov-sabelfeld:1995,sawford:2001}).  However, this
model does not require any Eulerian input and involves a
multiplicative noise ($\bm A$ depends on $\du$). Dimensional arguments
indicate that the local Kolmogorov time $\tau_\eta^\mathrm{loc}$ and
the acceleration amplitude $A \!=\! |\bm A|$ depends only on the
viscosity $\nu$ and on the local energy dissipation rate
$\epsilon_\mathrm{loc}$. We thus have $\tau_\eta^\mathrm{loc}\!\sim
\!\nu^{1/2}\, \epsilon_\mathrm{loc}^{-1/2}$ and $A \!\sim \!
\nu^{-1/4}\,\epsilon^{3/4}_\mathrm{loc}$.  These estimates predict
that the multiplicative term in (\ref{eq:model_da}) behaves as
$[\tau_\eta^\mathrm{loc}]^{1/2} A \!\sim\!
\epsilon_\mathrm{loc}^{1/2}$.  Interestingly this quantity is
independent on $\nu$ and is thus expected to have a finite limit at
infinite Reynolds numbers. Phenomenological arguments suggest that for
typical values of the velocity difference $\du$, the local dissipation
rate can be written as $\epsilon_\mathrm{loc} \sim [\delta
u^\parallel]^3/\delta x$, where $\delta u^\parallel =
\dx\cdot\du/\delta x$ is the longitudinal velocity difference between
the tracers. When $\delta u^\parallel=0$, the local dissipation rate
does not vanish but can be estimated through an averaged contribution
of larger eddies, leading to $\epsilon_\mathrm{loc} \simeq \epsilon$,
the averaged energy dissipation rate. These estimations have been
tested against numerical simulations: the inset of
Fig.~\ref{fig:autocor_accel} shows the variance of the acceleration
differences conditioned on the longitudinal velocity difference for
various separations. Up to some statistical errors, it seems that data
are in rather good agreement with the phenomenological prediction
which is shown as a dashed line. Finally such dimensional
considerations lead to model the large-time evolution of tracer
separation as
\begin{equation}
  \frac{\mathrm{d}\delta x}{\mathrm{d}t} = \delta u^\parallel, \quad
  \frac{\mathrm{d}\delta u^\parallel}{\mathrm{d}t} \sim
  \left[\epsilon + \alpha\,
    \frac{[\delta u^\parallel]^3}{\delta x}\right]^{1/2}\!\!\!\xi(t),
  \label{eq:model}
\end{equation}
where $\alpha$ is a positive parameter. Again here the multiplicative
noise is understood with Stratonovich convention. When rewriting it in
the It\=o sense, the additional drift that appears introduces a
``correlation time'' equal to the instantaneous turnover time $\delta
x/\delta u^\parallel$.  Preliminary studies of (\ref{eq:model}) showed
that its solutions follow a ballistic regime at short times and behave
according to Richardson law, i.e.\ $\langle\delta x^2\rangle\sim t^3$
at large times.  In this stochastic model, the local dissipation
$[\delta u^\parallel]^3/\delta x$ tends to a constant at large times,
so that in the asymptotic regime, the velocity difference obeys an
equation of the form ${\mathrm{d}\delta u^\parallel}/{\mathrm{d}t}
\propto \xi(t)$ and thus diffuses. So far, we have only investigated
the one-dimensional version (\ref{eq:model}) of the model. Extension
to higher dimensions requires accounting for incompressibility and is
the subject of ongoing work.

\begin{figure}[h]
  \includegraphics[width=\columnwidth]{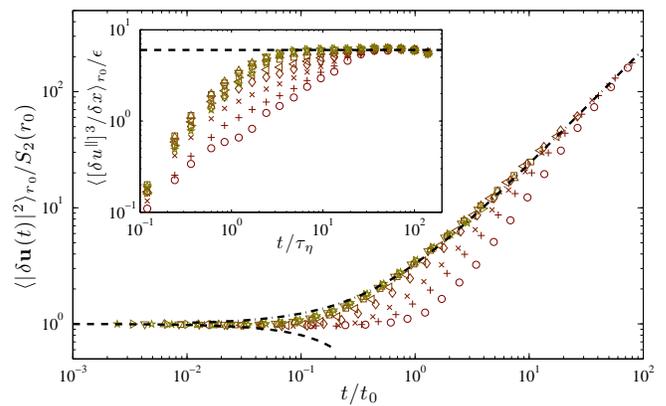}
  \caption{Time evolution of the averaged longitudinal velocity for
    $R_\lambda=730$ and different $r_0$ (same symbols as in
    Fig.~\ref{fig:batch_rich}). The short-time
    prediction~(\ref{eq:initsepvel}) is shown as a dashed line. The
    diffusive behavior $\langle |\du|^2\rangle_{r_0} \simeq
    S_2(r_0)+2.3\,\epsilon\,t$ is represented as a dash-dotted
    line. Inset: time evolution of $\langle[\delta u^\parallel]^3 /
    \delta x\rangle_{r_0}$; the dashed line corresponds to the value
    $6\epsilon$.}
  \label{fig:sepvel}
\end{figure}
To address the relevance of such a model to real flows, we turn back
to the analysis of simulation data. Figure \ref{fig:sepvel} shows the
time evolution of $\langle|\du(t)|^2\rangle_{r_0}$ for various values
of $r_0$. At small times this quantity slightly decreases because
the subleading term is negative. We indeed have $\du(t) \simeq
\du(0)+t\da(0)$, so that the ballistic regime reads
\begin{equation}
  \langle|\du(t)|^2\rangle_{r_0} = S_2(r_0)\,(1-2\,t/t_0) +\mathrm{h.o.t}.
  \label{eq:initsepvel}
\end{equation}
Again, the subleading terms are relevat for times
$t\lesssim0.01\,t_0$. Figure~\ref{fig:sepvel} also shows that at large
times the mean-squared velocity difference looses dependence on $r_0$
and grows $\propto\epsilon\, t$. In addition, as seen from the inset
of Fig.~\ref{fig:sepvel}, the averaged local dissipation rate
$\langle[\delta u^\parallel]^3/\delta x\rangle_{r_0}$ along particle
pairs approaches a positive constant $\simeq 6\,\epsilon$
(independently on $R_\lambda$) on times of the order of
$\tau_\eta$. This confirms the relevance of the mechanisms described
above in terms of a stochastic equation for the velocity differences.

Numerical results indicate that the time $t_0$ controls the
convergence to a diffusive regime for initial separations $r_0$ far
enough in the inertial range. This can be explained by the following
argument. As $\langle[\delta u^\parallel]^3/\delta x\rangle_{r_0}$
becomes constant on a short timescale, one expects that
\begin{equation}
\langle|\du(t)|^2\rangle_{r_0} \simeq S_2(r_0)+D\,\epsilon\,t \quad
\mbox{ for } t\gg\tau_\eta,
\end{equation}
where $D$ is a positive constant (for both Reynolds numbers, we
observe $D\approx 2.1$).  By balancing the diffusive term with the
initial mean-squared velocity difference $
\langle|\du(0)|^2\rangle_{r_0} = S_2(r_0)$, we find again that the
former is dominant for times $t$ much larger than $t_0$.  The
diffusive behavior of velocity differences is thus reached at times of
the order of $t_0$ and this explains in turn why this timescale
is that of convergence to Richardson's regime.

Let us summarize here our findings. In this work we give some evidence
that the Richardson explosive regime $\langle|\dx|^2\rangle \propto
t^3$ for the separation between two tracers in a turbulent flow
originates from a diffusive behavior of their velocity difference
rather than from dimensional arguments or equivalently a
scale-dependent eddy diffusivity for their distance. This leads on to
reinterpret the $t^3$ law as that of the integral of Brownian
motion. Such an argument is supported by two observations. First, the
acceleration difference has a short correlation time (of the order of
the Kolmogorov dissipative timescale) and can be approximated as a
white noise. Second, the amplitude of this noise solely depends on the
local dissipation rate $\langle [\delta u^\parallel]^3/\delta
x\rangle_{r_0}$, which becomes constant also on short
timescales. These considerations allow us to show that the time $t_0$
of convergence to Richardson's law is equal to that of deviations from
Batchelor's ballistic regime. This time, which reads $t_0 = S_2
(r_0)/(2\epsilon)$, is the time required to dissipate the kinetic
energy contained at a scale equal to the initial separation between
tracers.

The interpretation of Richardson's law as the diffusion of velocity
differences strongly questions possible effects of fluid-flow
intermittency on trajectory separation. Indeed, considerations on
velocity scaling, which are primordial in approaches based on eddy
diffusivity, are absent from the arguments leading to a diffusive
behavior of $\du$. Hence, we expect the separation $\dx$ to follow a
self-similar evolution in time, independently on the order of the
statistics. Intermittency will however affect
s directly the time of
convergence to such a regime. More frequent violent events (of tracer
pairs approaching or fleeing away in an anomalously strong manner)
will result in longer times for being absorbed by the average. Such
arguments do not rule out the possibility of having intermittency
corrections when interested in other observables than moments of the
separation, as it is for instance the case for exit times
\cite{boffetta-sokolov:2002b}. Such issues will certainly gain much
from a systematic study of multi-dimensional generalizations of the
stochastic model introduced here.

We ackowledge L.\ Biferale, G.\ Boffetta, M.\ Bourgoin, M.~Cencini,
G.~Eyink, G. Falkovich, A.\ Lanotte, E.~Villermaux for many useful
discussions and remarks.  Access to the IBM BlueGene/P computer JUGENE
at the FZ J\"ulich was made available through the ’XXL-project’
HBO28. The research leading to these results has received funding from
DFG-FOR1048 and from the European Research Council under the European
Community's Seventh Framework Program (FP7/2007-2013, Grant Agreement
no. 240579).

\end{document}